\documentclass[showpacs,preprintnumbers,amsmath,amssymb,twocolumn]{revtex4}

\usepackage{graphicx}
\usepackage{dcolumn}
\usepackage{bm}
\usepackage{color}
\begin{document}


\title{Particle Acceleration Around $5$-dimensional Kerr Black Hole}

\author{Ahmadjon Abdujabbarov$^{1,2}$}%
   \email{ahmadjon@astrin.uz }


\author{Naresh Dadhich$^{3,4}$}
\email{nkd@iucaa.ernet.in}

\author{Bobomurat Ahmedov$^{1,2,3}$}
 \email{ahmedov@astrin.uz}
 
 \author{Husan  Eshkuvatov$^{1,2}$}
  \email{husan@astrin.uz}

\affiliation{%
$^1$ Institute of Nuclear Physics, Ulughbek, Tashkent 100214, Uzbekistan\\
$^2$ Ulugh Beg Astronomical Institute,    Astronomicheskaya 33, Tashkent 100052, Uzbekistan\\
$^3$ Inter University Centre for Astronomy \& Astrophysics, Post
Bag 4, Pune 411007, India \\
$^4$ Centre for Theoretical Physics,
Jamia Millia Islamia, New Delhi 110025, India}
\date{\today}
\begin{abstract}
On the lines of the $4$-dimensional Kerr black hole we consider the
particle acceleration near a $5$-dimensional Kerr black hole which
has the two rotation parameters. It turns out that the center of
mass energy of the two equal mass colliding particles as expected
diverges for the extremal black hole and there is a symmetry in the
results for $\theta = 0, \pi/2$. Because of the two rotation
parameters, $r=0$ can be a horizon without being a curvature
singularity. It is shown that the acceleration of particles to high energies
near the 5-D extreme rotating black hole avoids fine-tuning of the
angular momentum of particles.

\end{abstract}

\maketitle

\section{\label{sec:intro}Introduction}

In the paper~\cite{banados09} 
authors have studied the 
center-of-mass 
(CM) energy of colliding 
two particles near the 
rotating black hole. 
They observed 
the divergence of the CM energy 
in the extreme rotating 
black hole case with the 
fine-tuning of the momentum 
of one of the particles 
(so called 
BSW process 
after the names of the authors - 
Banados, Silk, and West). 
The analysis of the 
CM energy 
of two colliding particles 
at the equatorial plane tends 
to extremely
high energies for the extremal 
central black hole when 
it rotates with
maximal speed 
and the maximal rotating 
black hole can be 
considered 
as high energy scale 
collider of 
normal and dark matter 
particles which 
can be detected by 
the observer at 
infinity. 
Since BSW process 
has been 
introduced 
the mechanism 
of particle acceleration 
near the black
hole has been 
intensively studied 
by many of authors 
for the different 
space-time 
metric describing 
black hole. 
Authors of the Ref.~\cite{gp11} have 
studied two particles acceleration, 
the multiple scattering and their 
CM energy in case of non extreme rotating black hole. 
The acceleration mechanism 
of the particles when 
they are in the stable 
circular orbits has 
been considered in~\cite{hk11}.

Recent studies have shown that the naked singularities that are formed
due to the gravitational collapse of massive stars provide a
suitable environment where particles could get accelerated and
collide at arbitrarily high center-of-mass energies~\cite{pj11,pj12,pj13,supspin}.

Authors of Ref.~\cite{Liu} studied the collision of two particles with the
different rest masses moving in the equatorial plane in a
Kerr-Taub-NUT spacetime and found that the CM energy depends not
only on the rotation parameter, but also on the NUT charge. 
The collision of particles in the vicinity of a horizon of a weakly
magnetized nonrotating black hole has been studied in \cite{frolov1}. 
Acceleration of particles by black hole with gravitomagnetic charge 
immersed in magnetic field~\cite{aa1}, by rotating black hole in 
a Randall-Sundrum brane with a cosmological constant~\cite{aa2}, 
and by rotating black hole in Ho\v{r}ava-Lifshitz gravity~\cite{aa3} 
have been studied in detail. Acceleration of electric current-carrying 
string loop near a Schwarzschild black hole embedded in 
an external magnetic field in the parallel direction 
to the axis of symmetry considered in~\cite{aa4}.   

Black holes are very interesting gravitational, as well as
geometric, objects which may exist in multidimensional spacetimes.
Other interesting axisymmetric object is the
five dimensional supergravity black hole \cite{chong}, which is an
impotant solution of supergravity Einstein-Maxwell equation.
Recently, a charged black hole solution in the limit of slow
rotation was constructed in \cite{Aliev06} ( also see
\cite{Aliev07}). Also, charged rotating black hole solutions have
been discussed in the context of supergravity and string theory
\cite[]{Cvetic1}-\cite[]{Cvetic05}. The solution obtained by Chong
\textit{et. al.} \cite{chong} of minimal gauged supergravity theory
comes closest to Kerr-Newman analogue. Energetics of a rotating
charged black hole in 5-dimensional supergravity spacetime has
been studied in~\cite{pd10} where energy
extraction even for axial fall has been predicted.

In this paper,
our main 
aim is to show 
particle acceleration 
for the axial
collisions by studying the 
collision of two particles 
with the same
rest masses in the 
background spacetime 
of the $5$-D Kerr black hole
and derive a general formula 
for the CM energy for the 
near-horizon
collision of two particles 
on the equatorial 
plane and polar plane.

The outline 
of the paper is the 
following. 
In the Sect.~\ref{motion} 
we
study the 
particle acceleration 
followed by the 
discussion of
particle collisions 
and the CM energy 
extraction
in the next 
Sect.~\ref{c.m.energy}. 
We
conclude 
with a discussion in the last 
Sect.~\ref{conclusion}.
Throughout the manuscript 
we use  $G=c=1$.

\section{\label{motion}
Particle Motion Around A Rotating Black Hole }

{The  Ricci}
{ flat metric for the 
5-dimensional Kerr
black hole in the 
Boyer- Lindquist coordinates 
$(t, r, \theta,
\varphi, \psi)$ has the following 
form~\cite{myers86}:}
\begin{eqnarray}\label{metric}
ds^2
&=&
-\frac{\Delta}{\rho^{2}}dT^{2}+\frac{\rho^{2}}
{\Delta}dr^{2}+
\rho^{2}d\theta^{2}+\rho^{2}\sin^{2}\theta
d\Phi^{2}\nonumber\\
&&+\rho^{2}\cos^{2}\theta
d\Psi^{2}
%
%
+\frac{\rho^{2}}{r^{2}}(b\sin^{2}\theta d\Phi
+a\cos^{2}\theta
d\Psi)^{2}\ ,\ \quad
\end{eqnarray}
where
\begin{eqnarray}
d T&=&
dt-a\sin^{2}\theta d\Phi-b\cos^{2}\theta d\Psi,\nonumber\\
d\nu&=& 
b\sin^{2}\theta d\Phi+a\cos^{2}\theta
d\Psi,\nonumber\\
\rho^{2}d\Phi&=& 
a dt-(r^{2}+a^{2})d\varphi,\nonumber\\
\rho^{2}d\Psi&=& 
b dt-(r^{2}+b^{2})d\psi,\nonumber\\
\Delta&=&
\frac{(r^{2}+a^{2})(r^{2}+b^{2})}{r^{2}}
-2M,\nonumber\\
\rho^{2}&=&
r^{2}+a^{2}\cos^{2}\theta+b^{2}\sin^{2}\theta\ .
\end{eqnarray}
Here $a$ and $b$ are the rotational parameters related to the
specific angular momenta of black hole with the total mass $M$
corresponding to the coordinates $\varphi$ and $\psi$, respectively.
The angular coordinates range over, $\theta \in[0,\pi/2]$ and
$\varphi,\psi \in [0,2\pi]$.

Using the equation $\Delta=0$ 
one can easily find the black hole horizon 
(higher positive root) 
in the form
\begin{eqnarray}
r_{+}&=&
\left[\left(M-\frac{a^{2}+b^{2}}{2}\right)+
\sqrt{\left(M-\frac{a^{2}+b^{2}}{2}\right)^{2}-a^{2}b^{2}}
\right]^{1/2},\nonumber
\end{eqnarray}
The horizon 
exists 
if 
$
a^{2}+b^{2}+ 2 |a||b| \leq 2M$.

The motion of particles and light in a space-time of a five-dimensional rotating black hole has been studied in~\cite{frolov2}. Complete integrability of geodesic motion in higher-dimensional rotating black-hole spacetimes has been studied in \cite{page1}.  
Here we will study 
the equation of 
motion for a test 
particle with
mass $m_{0}$ in 
the field of a 
$5$-dimensional 
rotating black hole.
The Lagrangian reads as
\begin{eqnarray} \label{lag}
{\cal L}&=& 
\frac{1}{2}g_{\mu\nu}\dot{x}^{\mu}\dot{x}^{\nu}
\end{eqnarray}
which 
readily leads to the conserved energy and angular momenta:

\begin{eqnarray}\label{elll}
-E= g_{tt}\dot{t}+g_{t\varphi}\dot{\varphi}+g_{t\psi}\dot{\psi}\ ,\\
\label{lphi}l_{\varphi}= g_{t\varphi}\dot{t}+ g_{\varphi\varphi}
\dot{\varphi}+g_{\varphi\psi}\dot{\psi}\ ,\\
\label{lpsi} l_{\psi}= g_{t\psi} \dot{t}+g_{\varphi\psi}
\dot{\varphi}+g_{\psi\psi}\dot{\psi}\ .
\end{eqnarray}
Solving the equations (\ref{elll}) -- (\ref{lpsi}), one can write
\begin{eqnarray}
\frac{dt}{ds}&=&
-\Upsilon^{-1}\big[E (g_{\varphi\psi}^2 -
g_{\varphi\varphi} g_{\psi\psi}) - l_{\psi}g_{\varphi\varphi} g_{
t\psi}
\nonumber\\&&
+ (l_{\psi} g_{t\varphi} + l_{\varphi}
g_{t\psi})g_{\varphi\psi}- l_{\varphi} g_{\psi\psi}g_{ t\varphi}
\big]\ ,\\
\frac{d\varphi}{ds}&=&
-\Upsilon^{-1}\big[E (g_{\psi\psi}
g_{t\varphi} -  g_{\psi\varphi} g_{t\psi}) + (l_{\psi}
g_{t\varphi}  - l_{\varphi} g_{t\psi}) g_{t\psi}\nonumber\\&& -
(l_{\psi}
g_{\psi\varphi} -l_{\varphi} g_{\psi\psi}) g_{tt}\big]\ ,\\
\frac{d\psi}{ds}&=&
-\Upsilon^{-1}\big[ E( g_{\varphi\varphi}
g_{t\psi}- g_{\psi\varphi} g_{t\varphi}) - (l_{\psi} g_{t\varphi}
- l_{\varphi} g_{t\psi}) g_{t\varphi}\nonumber\\&&  + (l_{\psi}
g_{\varphi\varphi}  - l_{\varphi} f g_{\psi\varphi}) g_{tt}\big]\ ,
\end{eqnarray}
where
\begin{eqnarray}
\Upsilon&=&{( g_{\psi\psi} g_{t\varphi}^2 - 2 g_{\psi\varphi}
g_{t\varphi} g_{t\psi} + g_{\varphi\varphi} g_{t\psi}^2 +
g_{\psi\varphi}^2 g_{tt} - g_{\varphi\varphi} g_{\psi\psi}
g_{tt})}\ .\nonumber
\end{eqnarray}

The metric functions have the following form:
\begin{eqnarray}
g_{tt}&=&1-\frac{2 M }{\rho^2}\nonumber\ ,\\
g_{t\varphi}&=&-\frac{2aM}{\rho^2}\sin^2\theta\nonumber\ ,\\
g_{t\psi}&=&-\frac{2bM}{\rho^2}\cos^2\theta\nonumber\ ,\\
g_{\varphi\varphi}&=&(r^2+a^2)\sin^2\theta+\frac{2aM}{\rho^2}a\sin^4\theta\nonumber\ ,\\
g_{\psi\psi}&=&(r^2+b^2)\cos^2\theta+\frac{2bM}{\rho^2}b\cos
^4\theta\nonumber\ ,\\
g_{\varphi\psi}&=&\frac{2abM}{\rho^2}\sin^2\theta\cos^2\theta\nonumber\ ,\\
g_{rr}&=&\frac{\rho^2}{\Delta}\ ,\qquad
g_{\theta\theta}=\rho^2\nonumber\ .
\end{eqnarray}

Now for the motion in the polar plane $\theta=0$, we have
$l_{\varphi}=0,~ \rho_a^{2}=r^{2}+a^{2}$ and $\dot{\theta}=0,$
\begin{eqnarray}
\label{t}
\frac{dt}{d\tau}&=&\frac{E}{(\rho_a^{2}-2
M)}\left[\rho_a^{2}-\frac{
4M^{2}b^{2}}{b^{2}{\rho_a^{2}} +r^{2}(\rho_a^{2}-2M)}\right]\nonumber\\
&& -L_{\psi}\frac{2M b}{b^{2}{\rho_a^{2}}+r^{2}(\rho_a^{2}-2M)},\\ \nonumber\\
\label{psi}
\frac{d\psi}{d\tau}&=&\frac{(\rho_a^{2}-2M)L_{\psi}+2 b M E
}{b^{2}\rho_a^{2} +r^{2}(\rho_a^{2}-2M)},\\ \nonumber\\
\label{r}
\left(\frac{dr}{d\tau}\right)^2&=&\frac{\Delta}{\rho_a^{2}}
\bigg\{\frac{E^{2}}{\rho_a^{2}-2M}-1\nonumber\\
&&-\frac{\big[(\rho_a^{2}-2 M) L_{\psi} +2 b M E
]^{2}}{(\rho_a^{2}-2M)\left[b^{2}\rho_a^{2}
+r^{2}(\rho_a^{2}-2M)\right]}\bigg\}   .
\end{eqnarray}
%
%

Note that the motion in 
the equatorial plane,  
$\theta=\pi/2$,
will be given 
by letting 
$\psi\to\phi, a\to b, b\to a$.

Now one 
can easily write 
the radial equation of 
motion for the test massive 
particle in the equatorial 
plane and polar 
plane in the 
following form
\begin{eqnarray}
\frac{1}{2}\dot{r}^{2} + V_{\rm eff}(r)&=&0,
\end{eqnarray}
where the quantity 
\begin{eqnarray}
V_{eff}(r)&=&\frac{1}{2}\frac{\Delta}{\rho_a^{2}}\bigg\{
1-\frac{E^{2}}{\rho_a^{2}-2M}\nonumber\\
&&+\frac{\big[(\rho_a^{2}-2 M) L_{\psi} +2 b M E
]^{2}}{(\rho_a^{2}-2M)\left[b^{2}\rho_a^{2}
+r^{2}(\rho_a^{2}-2M)\right]}\bigg\}\ ,\ \
\end{eqnarray}
%
can be interpreted 
as an effective 
potential for 
the polar plane 
$\theta=0$
and similar form for 
the effective 
potential  at 
equatorial plane
($\theta=\pi/2$) 
with transformations 
$a\to b$ 
and 
$b \to a$.

To consider 
the circular orbits 
one should use 
the following conditions:
\begin{eqnarray} \label{orbit}
&& V_{eff}(r)=0, ~~~\frac{d V_{eff}(r)}{d r}=0,
\end{eqnarray}
This leads to a limitation 
on the possible values 
of the angular
momentum for 
collision of two 
particles and 
after some
straightforward 
calculation, one can 
obtain the range 
of angular
momenta of particles 
for the special cases 
when $a=b$:
\begin{eqnarray}  \label{range}
\frac{-a-\sqrt{-a^{2}+2a^{4}}}{a^{2}-1}\leq l\leq
\frac{3a+\sqrt{a^{2}-1} }{2},
\end{eqnarray}
and when $a=-b$
\begin{eqnarray}  \label{range1}
-\frac{3a+\sqrt{a^{2}-1} }{2}\leq
l\leq\frac{a+\sqrt{-a^{2}+2a^{4}}}{a^{2}-1}.
\end{eqnarray}
One should have 
the satisfied 
condition 
in the polar plane
\begin{eqnarray}
\label{ineq} E \left(\rho_a^{4}(r^{2}+b^{2})+b^{2}f_a \right) \geq
L bf_a \ ,
\end{eqnarray}
in order to 
have 
${dt}/{d\tau}\geq 0$. 

In the limiting case 
when $r \rightarrow r_{+}$ 
for the massive test
particle, one can gets
\begin{eqnarray}
E \geq \frac{bf_a  }{\rho_a^{4}(r^{2}+b^{2})+b^{2}f_a}L=\omega_{H}
L.\nonumber
\end{eqnarray}
\begin{figure}
\includegraphics[width=0.45\textwidth]{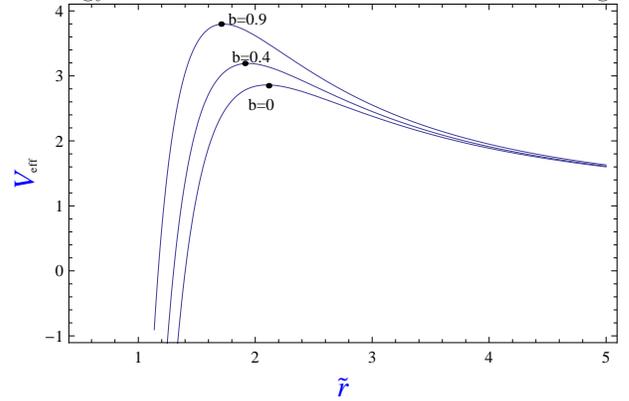}

\caption{\label{fig:1} The 
radial dependence 
of the effective
potential of the 
radial motion 
of the test particles 
in the polar
plane, ($\theta=0$), 
for different values 
of the parameter $b$.
For the motion 
in the equatorial 
plane, $\theta=\pi /2$, 
with
$b\to a$, 
the graphs will be exactly 
the same.}
\end{figure}

\section{\label{c.m.energy}
Center-Of-Mass Energy For A  Rotating Black Hole In 5 Dimensional
Spacetime}

This section 
is devoted to study 
the CM energy
of the accelerating 
particles near the 
rotating black
hole in 5-dimensional 
spacetime. 
Hereafter 
we assume 
that the
motion of 
particles 
occurs both 
in the equatorial 
plane
and the polar  
of a rotating 
black hole. 
For simplicity 
one may consider 
that two
colliding particles 
has the same 
rest mass 
$m_{0}$ and 
they are at 
rest at
infinity ($E=m_{0}$), 
then they 
approach 
the rotating black hole
and collide at some radius $r$. 
We assume that two particles $1$
and $2$ 
are at the 
same spacetime 
position and have 
angular momenta
$l_{1}$ and $l_{2}$, 
respectively. 
Here, our aim is 
to compute the
energy in the 
CM frame for this collision 
according to
the calculation method 
developed in~\cite{banados09}. 
The 
momentum of the particle $i$ ($i=1,2$) 
is given by
\begin{eqnarray}
p_{i}^{\mu}=m_{0}u_{i}^{\mu}, \nonumber
\end{eqnarray}
where $u_{i}^{\mu}$ is 
the velocity of particles $i$. 
One can construct the total momentum of two particles in the form
\begin{eqnarray}
p_{t}^{\mu}&=&p_{1}^{\mu}
+p_{2}^{\mu}. \nonumber
\end{eqnarray}
It is easy to calculate the CM energy $E_{c.m}$ of accelerating particles using the standard formulae
\begin{eqnarray}\label{c.m.}
E_{c.m.}
&=&
\sqrt{2}m_{0}\sqrt{1-g_{\mu\nu}u^{\mu}_{(1)}u^{\nu}_{(2)}}\ .
\end{eqnarray}

Here, 
we consider 
two particles 
coming from 
infinity with
$E_{1}/m_{0}=E_{2}/m_{0}=1$ 
for simplicity. 
Inserting equations
(\ref{t})--(\ref{r}) 
and equations of 
motion at the 
equatorial plane
into the expression (\ref{c.m.}), 
one can easily  
calculate CM energies 
of the accelerating 
particles near the 
rotating black hole in
the two cases of the 
5 dimensional spacetime.

In the first case, 
specializing to 
motion along 
$\theta=0$, we
have $L_{\varphi}=0$ 
and $g_a=\rho_a^4-f_a$. 
CM energy of the two
particles is calculated as
\begin{widetext}
\begin{eqnarray}
\label{c.m.1} 
&&\frac{E^{2}_{\rm c.m.}}{2m_{0}^{2}}  =  1 +
\frac{\rho_a^{2}}{\rho_a^{2}-2M} - \left[\frac{4b^{2}
M^{2}}{\rho_a^{2} (\rho_a^{2}-2M)} +r^{2}+b^{2}+\frac{b^{2}
(\rho_a^{2}-2M)}{\rho_a^{2}}\right]
\frac{\left[l_{1}(\rho_a^{2}-2M)
+2bM\right]\left[l_{2}(\rho_a^{2}-2M)+2b
M\right]}{\left[b^{2}\rho_a^{2}
+r^{2}(\rho_a^{2}-2M)\right]^{2}} \nonumber\\
&&-\frac{1}{(\rho_a^{2}-2M) \left[b^{2}\rho_a^{2}+r^{2}(\rho_a^{2}-2M)\right]} \nonumber\\
\\
&&\times \sqrt{ \left\{
2M\left[b^{2}\rho_a^{2}+r^{2}(\rho_a^{2}-2M)\right]-\left[l_{1}(\rho_a^{2}-2M)
+2bM\right]^{2}\right\}{\left\{
2M\left[b^{2}\rho_a^{2}+r^{2}(\rho_a^{2}-2M)\right]-\left[l_{2}(\rho_a^{2}-2M)
+2bM\right]^{2}\right\}}} \ \nonumber,
\end{eqnarray}
\end{widetext}
and similar form for the 
expression for CM 
energy of the two
particles  at 
equatorial plane
($\theta=\pi/2$) with 
transformations 
$a\to b$ and 
$b \to a$. 

%
%
%
%
%
%
%

\subsection{Classification of center of mass energy of two colliding particles near rotating 5 dimensional black hole}

\begin{figure}
\includegraphics[width=0.45\textwidth]{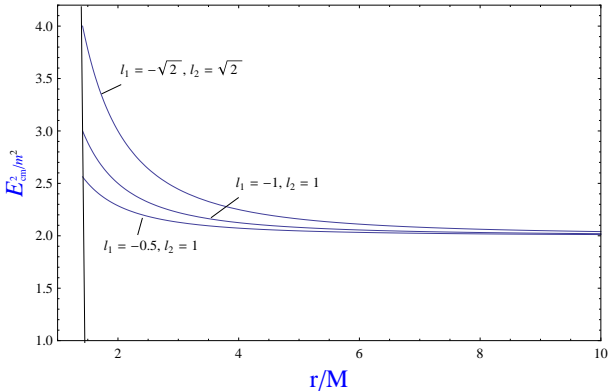}
\caption{\label{fig00} Radial 
dependence of 
the center of mass
energy of accelerating 
particles around rotating 
five dimensional
black hole for the 
different values of 
the angular 
momentum of the
particles in the 
case when $a=b=0$. 
The vertical line 
corresponds to 
event horizon.}
\end{figure}

\begin{figure}
\includegraphics[width=0.45\textwidth]{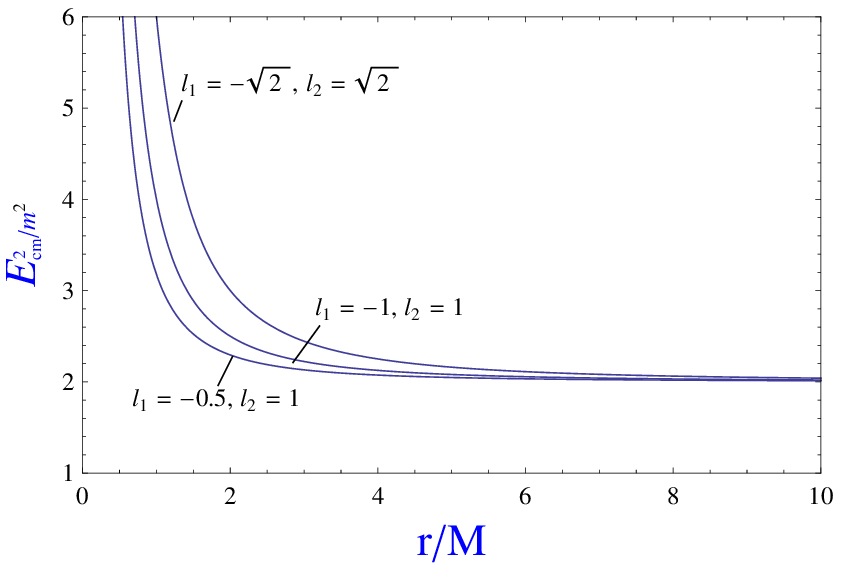}
\caption{\label{fig01} Radial 
dependence of the 
center of mass
energy of accelerating 
particles around rotating 
five dimensional
black hole for the different 
values of the angular 
momentum of the
particles in the case 
when $b=0$, $a=\pm\sqrt{2}$ 
which is
relevant to extremal 
rotating black hole. 
The value of the 
event horizon 
radius is $r_{+}=0$.}
\end{figure}

\begin{figure*}
a) \includegraphics[width=0.45\textwidth]{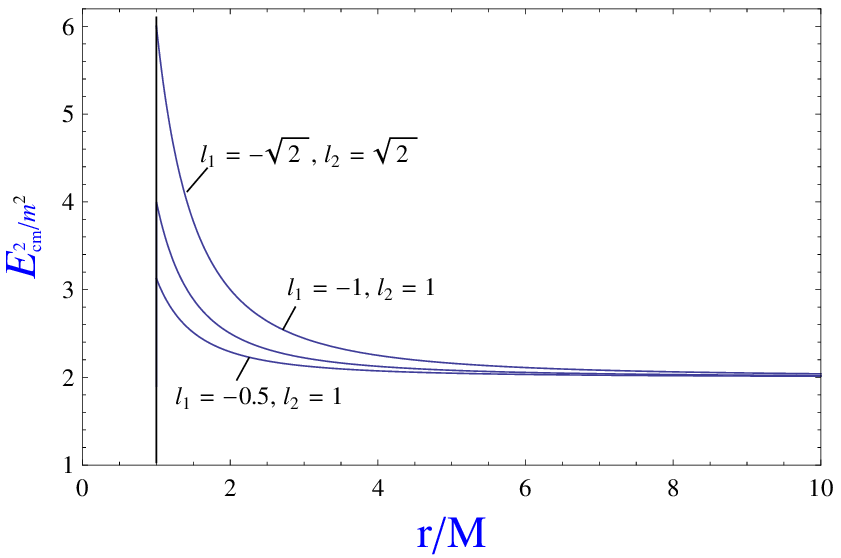} b)
\includegraphics[width=0.45\textwidth]{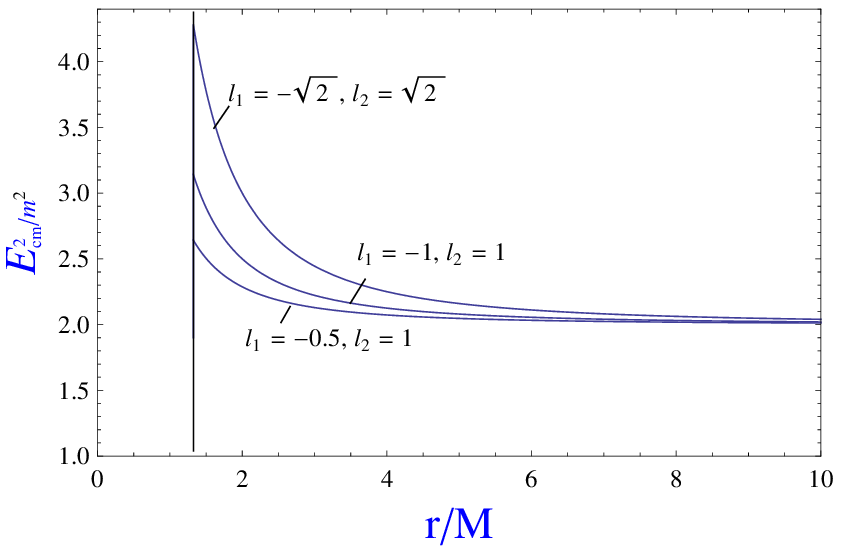}
\caption{\label{fig023} 
Radial dependence 
of the center of mass
energy of accelerating 
particles around 
rotating five dimensional
black hole for the 
different values 
of the angular 
momentum of the
particles in the cases 
when a) $b=0$, $a=1$ 
and b) $b=0$, $a=0.5$
which are  relevant 
to nonextremal rotating 
black hole. The vertical 
lines correspond to 
the event horizon. }
\end{figure*}

\begin{figure}
\includegraphics[width=0.45\textwidth]{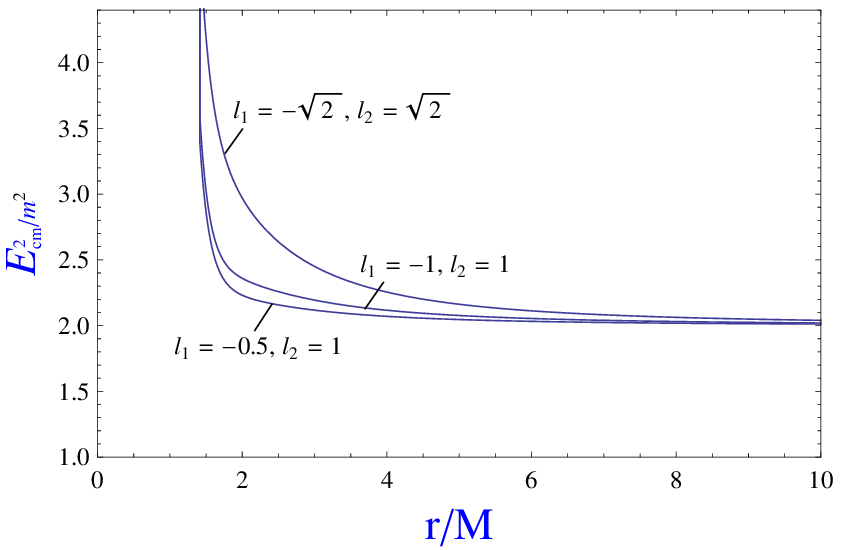}
\caption{\label{fig5} Radial 
dependence of the center 
of mass
energy of accelerating 
particles around 
rotating five dimensional
black hole for the 
different values of 
the angular 
momentum of the
particles in the case 
when $a=0$, $b=\sqrt{2}$ 
which is
corresponding to 
extremal rotating 
black hole. 
The value of the 
event horizon 
radius is $r_{+}=0$.}
\end{figure}
\begin{figure}
\includegraphics[width=0.45\textwidth]{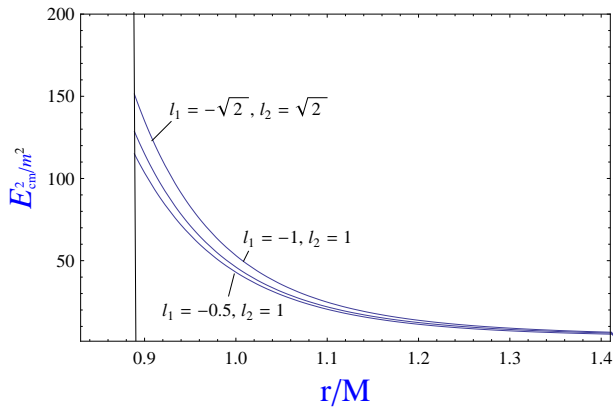}
\caption{\label{fig6} 
Radial dependence 
of the center of mass
energy of accelerating 
particles around 
rotating five dimensional
black hole for the 
different values of 
the angular 
momentum of the
particles in nonextremal   
case of $b=0$, $b=1.1$. 
The vertical line 
corresponds to 
the event horizon.}
\end{figure}
\begin{figure*}
\includegraphics[width=0.32\textwidth]{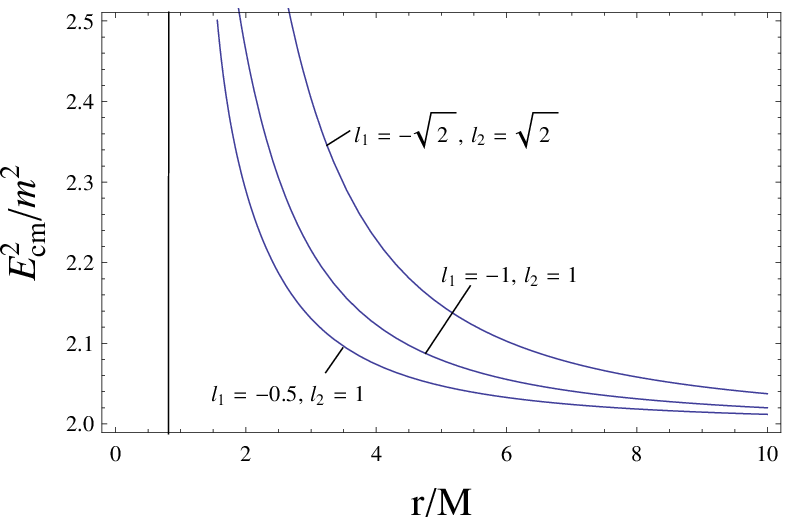}
\includegraphics[width=0.32\textwidth]{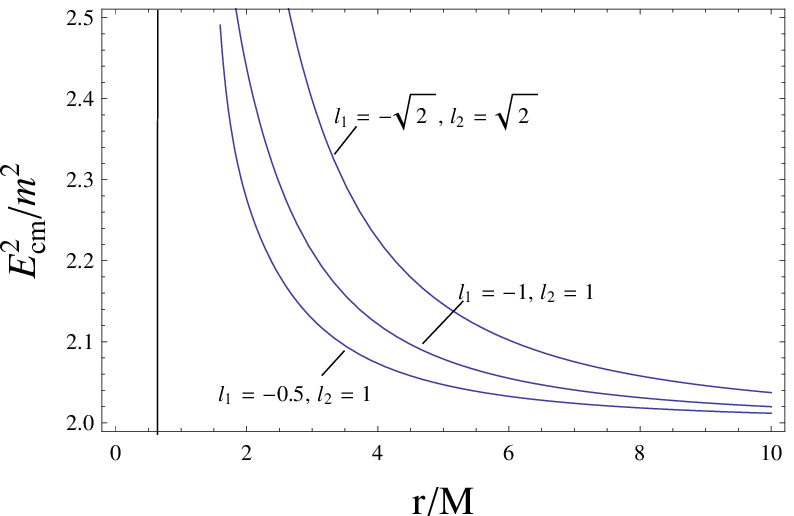}
\includegraphics[width=0.32\textwidth]{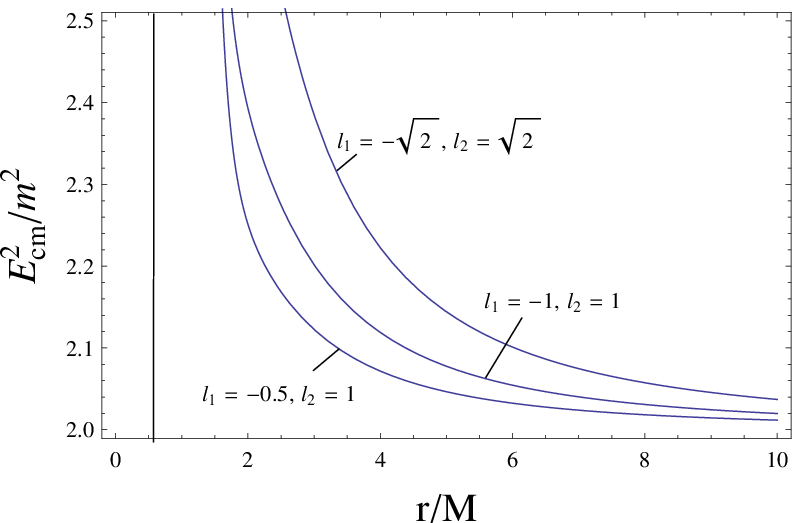}

\includegraphics[width=0.32\textwidth]{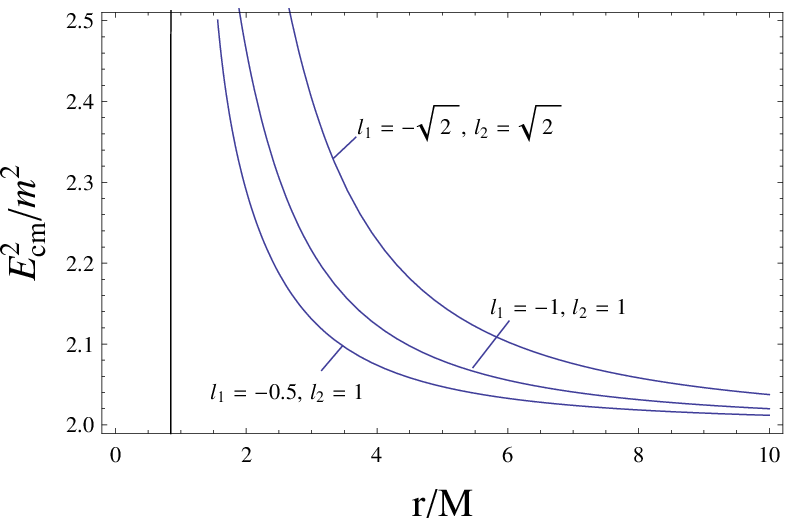}
\includegraphics[width=0.32\textwidth]{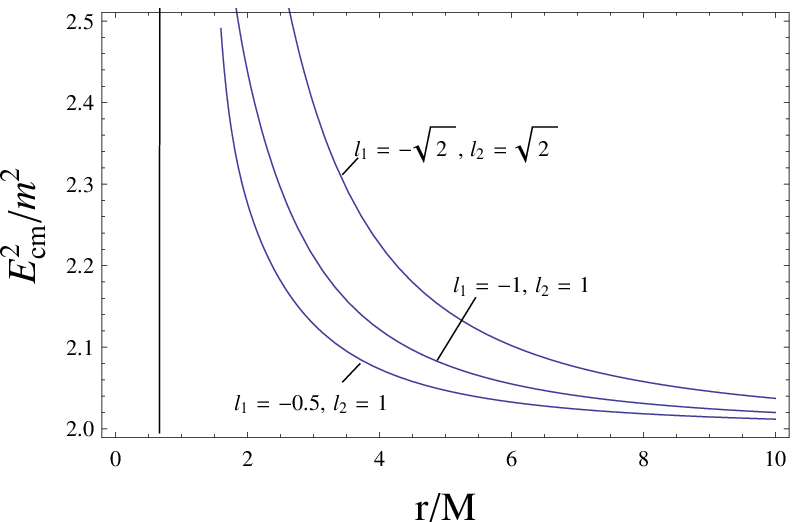}
\includegraphics[width=0.32\textwidth]{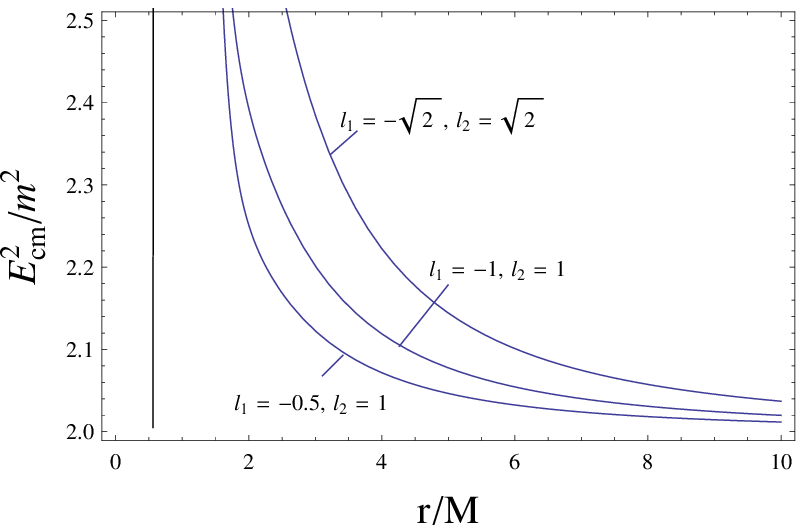}

\caption{\label{fig7} Radial 
dependence of the 
center of mass
energy of accelerating 
particles around 
rotating five dimensional
black hole for the different 
values of the angular 
momentum of the
particles in extremal  
case when $a=\sqrt{2}-b$  
(upper plots) and  $a=b-\sqrt{2}$ 
(lower plots). From left to the right the 
plots correspond to the 
case when $b=0.7$, $b=0.9$, 
5
and $b=1.2$, respectively. 
5
The vertical line corresponds to the event horizon.}
\end{figure*}
{Below we will analyze the expression for CM energy of two particles  (\ref{c.m.1}). In all cases
mass of BH is taken to be $M=1$.}

\begin{itemize}

\item If both rotational parameters
are vanishing: $ a=b=0$, then event horizon is located at
$r_{+}=\sqrt{2}$. Center of mass energy is finite in this case and
has the following limit:
\begin{eqnarray}
\frac{E_{\rm c.m.}^2}{m^2}=(l_{1}-l_{2})^2+8\ .
\end{eqnarray}

The radial dependence of the center of mass energy for the
different values of the angular momentum of the particles are
shown in the Fig.~\ref{fig00}

\item Rotational parameter $b$ is  vanishing:
$b=0$. If condition $a=\pm \sqrt{2}$ will be satisfied, then BH is
extreme. Center of energy diverges in any values of angular
momentums of the particles in the range : $-\sqrt{2}\leq l_{1}\leq \sqrt{2}$,
$-\sqrt{2}\leq l_{2}\leq \sqrt{2}$:
\begin{eqnarray}
\frac{E_{\rm c.m}^{2}}{2m^2}=2+\frac{2-l_{1}l_{2}-\sqrt{(2
-l_{1}^{2})(2-l_{2}^{2})}}{r^2}\ .
\end{eqnarray}

The radial dependence of the center of mass energy for the
different values of the angular momentum of the particles are
shown in the Fig.~\ref{fig01}. From this dependence one 
may observe that with the increasing the module of the 
expression $(l_{1}-l_{2})$ the center of mass energy 
tends to the higher values more faster for infalling particles.

\item Rotational parameter $b$ is vanishing:
$b=0$. If condition $a^2<2$ will be satisfied, then BH is nonextreme
and horizon located at $0<r_{+}\leq \sqrt{2}$. Center of mass
energy is finite. The radial dependence of the center of mass
energy of the particles in the different values of the angular
momentum of the particle are shown in the Fig.~\ref{fig023} when
a) $a=1$, $r_{+}=1$ and b) $a=0.5$, $r_{+}=\sqrt{7/4}$. Note that
in particular case when $a=1$ the center of mass energy has the following form:
\begin{eqnarray}
\frac{E_{\rm c.m.}^2}{m^2}=\frac12\big[(l_{1}-l_{2})^2+4\big]\ .
\end{eqnarray}

\item Rotational parameter $a$ is vanishing:
$a=0$. If the condition $b=\pm \sqrt{2}$ will be satisfied, BH is
extreme. Center of  energy diverges in any values of angular
momentums of the particle: $-\sqrt{2}\leq l_{1}\leq \sqrt{2}$,
$-\sqrt{2}\leq l_{2}\leq \sqrt{2}$. The radial dependence of the center of mass
energy of the particles in the different values of the angular
momentum of the particle are shown in the Fig.~\ref{fig5}.

\item Rotational parameter $a$ is vanishing:
$a=0$.  If condition  $b^2<2$, $b^2\neq 1$ will be satisfied, center of
mass is finite. The radial dependence of the center of mass
energy of the particles for the different values of the angular
momentum of the particle are shown in the Fig.~\ref{fig6}.

\item  The
following condition should be satisfied to be extremal BH: (i)
$1-(a+b)^2/2=0$ and (ii) $1-(a-b)^2/2=0$. let us consider extra
condition: $r_{+}=0\ \ \Rightarrow \ \ a^2+b^2=2$ then one can
find the solution for $a$ and $b$ as : $a=0, \ b=\pm \sqrt{2}$ and
as: $b=0, \ a=\pm \sqrt{2}$. In all cases the center of mass
energy diverges.

\item {Consider the extremal rotating 5-D black hole with nonvanishing $r_{+}$. This implies the conditions i) $a=\sqrt{2}-b$ and ii) $a=b-\sqrt{2}$. In Fig.~\ref{fig7} the radial dependence of the center of mass
energy of the particles for the different values of the angular
momentum of the particle are shown. The upper and lower plots correspond to the condition (i) and (ii), respectively. From this dependence one may conclude that the centre of mass energy of the particles diverge near the event horizon when the central object is the 5-D extreme rotating black hole. However, one can see that the fine-tuning for the angular momentum of the particles is not required for 5-D rotating black hole. Note that for 4 dimensional rotating black hole one needs significant fine-tuning to get sensible cross sections
for particles. 
}
\end{itemize}

\section{\label{conclusion} Conclusion}

The study 
described in this 
manuscript devoted to 
particles acceleration 
mechanism
at the equatorial plane 
and polar region of a
rotating black hole 
in the 5 dimensional  
spacetime. 
We have derived
a general formula 
for the CM energy 
and made analyse  it
for the different cases. 
It was pointed out by the Banados,
Silk, and West~\cite{banados09} 
that a rotating black hole 
in 4 dimensional 
spacetime can, 
in principle,
accelerate the 
particles falling to the 
central black
hole to arbitrarily 
high energies if 
one of the particles has
angular momentum $\ell=2$.
We have
derived a general 
formula for the CM 
energy near the horizon on
the equatorial plane and polar plane.

We have found that particles will 
collide near-extremal singularity  
and  the center of mass
energy for collision of the two 
particles can be unlimited
near-extremal singularity of the 
5 dimensional  spacetime. 
{Our result shows
that arbitrarily high 
CM energy appears near-extremal singularity
even for the axial collision which 
is a significant difference
from other black holes. 
In particular, energy could be extracted
even in the polar  region through 
$aq$ coupling producing a
rotation.  This is similar to the 
energy extraction by Penrose
process discussed in 
the paper \cite{pd10} by 
one of the authors of this 
paper. }

The frame-dragging effects 
in a pure Kerr black hole spacetime
can accelerate particles 
and sone needs significant 
fine-tuning to get sensible 
cross sections
for particles (at least one 
of particles has to have 
critical angular momentum). 
Recently it was shown that 
the acceleration process near 
the 4 dimensional naked 
singularity avoids fine-tuning 
of the parameters of the particle 
geodesics for the unbound center 
of mass energy of collisions~\cite{pj11,pj12,pj13,supspin}. 
Here we show that the CM 
energy diverge for any 
values of particles 
falling to central objects. 
The unbound CM 
energy can be observed for 
any particles coming 
inward to 5 dimensional 
extreme rotating black hole.
%


%
\begin{acknowledgments}

A.~A. and B.~A. thank the TIFR, IUCAA (India), and Faculty of
Philosophy and Science, Silesian University in Opava
(Czech Republic) for warm
hospitality. This research is supported in part by the
projects F2-FA-F113, FE2-FA-F134, and F2-FA-F029 of
the UzAS and by the ICTP through the OEA-PRJ-29
project. A.~A. and B.~A. acknowledge the German
Academic Exchange Service (DAAD), the Volkswagen�
Stiftung and the TWAS Associateship grants, and thank the
Max Planck Institut f\"{u}r GravitationsPhysik, Potsdam for
the hospitality.

\end{acknowledgments}

\end{document}